\documentclass[reqno]{article}
\newcommand{\ds}{\displaystyle}
\newcommand{\dsf}{\ds\frac}

\newcommand{\beq}{\begin{equation}}
\newcommand{\eeq}{\end{equation}}

\usepackage{ae} 
\usepackage[T1]{fontenc}
\usepackage[ansinew]{inputenc}
\usepackage{amsmath}
\usepackage{graphicx}
\usepackage{color}
\usepackage[colorlinks]{hyperref}
\usepackage{multicol}
\begin{document}
\small

\begin{center}
\bf Branching instability in the flux creep regime of type-II
superconductors
\end{center}
\begin{center}
N. A. Taylanov
\end{center}
\begin{center}
\emph{\footnotesize National University of Uzbekistan}
\end{center}
\begin{center}
 \textcolor{black}{Abstract}
\end{center}
\begin{center}
\mbox{\parbox{13cm}{\footnotesize  We study theoretically the
space-time evolution of the thermal and electromagnetic
perturbation in a superconductor with a nonlinear current-voltage
characteristics in the flux creep regime. On the basis of a linear
analysis of a set of differential equations describing small
perturbations of temperature and electromagnetic field, it is
found that under some conditions a branching instability may occur
in a superconductor sample.}}
\end{center}

\begin{multicols}{2}{The dynamics of
thermomagnetic instabilities of the critical state and flux jumps
in hard superconductors with high values of the critical current
density and the critical magnetic fields have been earlier
investigated by many authors [1-5]. The general concept of
thermomagnetic instabilities in type-II superconductors was
developed in literature [5]. The dynamics of small thermal and
electromagnetic perturbations, whose development leads to the flux
jump, have been investigated theoretically by Mints and Rakhmanov
[4, 5]. The authors have found the stability criterion for the
flux jumps in the framework of adiabatic and dynamic
approximations in the viscous flux flow regime of type-II
superconductors. The detailed theoretical analyze of the flux
jumping in the flux creep regime, where the current-voltage
characteristics of a sample is a nonlinear have been carried out
recently by Mints [6] and by Mints and Brandt [7]. Nonuniform
magnetic flux penetration in type II superconductors, creating
finger and dendritic patterns, has recently attracted considerable
interest. Such patterns have been directly observed in a large
number of superconducting films employing magneto-optical imaging
techniques [8-15]. The existing experimental data [8-15], and the
recently developed theoretical models [16, 17], suggest that the
origin of these patterns is thermomagnetic instability of the
vortex matter in the superconducting films. The instability arises
from local temperature increase due to flux motion, which, in
turn, decreases flux pinning and hence facilitates further flux
motion. Linear stability analysis, based on the coupled nonlinear
Maxwell and thermal diffusion equations, showed that instability
in the form of narrow fingers perpendicular to the background
electric field occurs when this field exceeds of its threshold
value  [16, 17].

In the present work, we study the spatial and temporal evolution
of small thermal and electromagnetic perturbation in type-II
superconductor sample in the flux creep regime with a nonlinear
voltage-current characteristics, assuming that an applied field
parallel to the surface of the sample. On the basis of a linear
analysis of a set of differential equations describing small
perturbations of temperature and electromagnetic field we found
that under some conditions a branching instability may occur in
the sample.

Mathematical problem of theoretical study the dynamics of thermal
and electromagnetic perturbations in a superconductor sample in
the flux creep regime can be formulated on the basis of a system
nonlinear diffusion-like equations for the thermal and
electromagnetic field perturbations with account nonlinear
relationship between the field and current in superconductor
sample. The distribution of the magnetic flux density $\vec B$ and
the transport current density $\vec j$ inside a superconductor is
described by the equation

\begin{equation}
rot\vec B=\mu_0\vec j.
\end{equation}
When the penetrated magnetic flux changes with time, an electric
field $\vec E$ is generated inside the sample according to
Faraday's law
\begin{equation}
rot\vec E=\frac{d\vec B}{dt}.
\end{equation}
The temperature distribution in superconductor is governed by the
heat conduction diffusion equation

\begin{equation}
\nu (T)\dsf{dT}{dt}=\nabla[\kappa(T)\nabla T]+\vec j\vec E,
\end{equation}

Here $\nu=\nu(T)$ and $\kappa=\kappa(T)$ are the specific heat and
thermal conductivity, respectively. The above equations should be
supplemented by a current-voltage characteristics of
superconductors, which has the form
$$
j=j_{c}(T,B)+j(E).
$$
In order to obtain analytical results of a set Eqs. (1)-(3), we
suggest that $j_c$ is independent on magnetic field induction $B$
and use the Bean critical state model $j_c=j_c(B_e,
T)=j_0-a(T-T_0)$ [1, 18], where $B_e$ is the external applied
magnetic field induction; $a=j_0/(T_c-T_0)$; $j_0$ is the
equilibrium current density, $T_0$ and $T_c$ are the equilibrium
and critical temperatures of the sample, respectively, [5]. For
the sake of simplifying of the calculations, we perform our
calculations on the assumption of negligibly small heating
$(T-T_0\ll T_c-T_0)$ and assume that the temperature profile is a
constant within the across sample and thermal conductivity
$\kappa$ and heat capacity $\nu$ are independent on the
temperature profile.  We shall study the problem in the framework
of a macroscopic approach, in which all lengths scales are larger
than the flux-line spacing; thus, the superconductor is considered
as an uniform medium. The system of differential equations (1)-(3)
should be supplemented by a current- voltage curve $j=j(E)$. In
the flux creep regime the current-voltage characteristics of type
II conventional superconductors is highly nonlinear due to
thermally activated dissipative flux motion [19, 20]. For the
logarithmic current dependence of the potential barrier $U(j)$,
proposed by [21] the dependence $j(E)$ has the form

\begin{equation}
j=j_c\left[\dsf{E}{E_0}\right]^{1/n}.
\end{equation}
where $E_0$ is the voltage criterion at which the critical current
density $j_c$ is determined [5]; a constant parameter n depends on
the pinning regimes and can vary widely for various types of
superconductors. In the case $n=1$ the power-law relation (4)
reduces to Ohm's law, describing the normal or flux-flow regime
[18]. For infinitely large $n$, the equation describes the Bean
critical state model $j=j_c$ [1]. When $1<n<\infty$, the equation
(4) describes nonlinear flux creep [22]. In this case the
differential conductivity $\sigma$ is determined by the following
expression

\begin{equation}
\sigma=\dsf{d\vec j}{d\vec E}=\dsf{j_c}{nE_b}
\end{equation}
According to relation (5) the differential conductivity decreases
with the increasing of  the background electric field $E_b$, and
strongly depends on the external magnetic field sweep rate
$E_b\sim \dot B_ex$. Therefore the stability criterion also
strongly depends on the differential conductivity $\sigma$. For
the typical values of $j_1=10^3 A/cm^2$, $E_b=10^{-7} V/cm$ we
obtain $\sigma=10^{10} 1/\Omega cm$. It follows from this
estimation [6, 7] that the differential conductivity $\sigma$
which determines the dynamics of thermomagnetic instability for is
high enough. We assume, for simplicity, that the value of n
temperature and magnetic-field independent.

Let us formulate a differential equations governing the dynamics
of small temperature and electromagnetic field perturbation in a
superconductor sample. We study the evolution of the thermal and
electromagnetic penetration process in a simple geometry -
superconducting semi-infinitive sample $x\geq 0$. We assume that
the external magnetic field induction $B_e$ is parallel to the
z-axis and the magnetic field sweep rate $\dot{B_e}$ is constant.
When the magnetic field with the flux density $B_e$ is applied in
the direction of the z-axis, the transport current $j(x, t)$ and
the electric field $E(x, t)$ are induced inside the slab along the
y-axis. For this geometry the spatial and temporal evolution of
small thermal T(x, t) and electromagnetic field E(x, t)
perturbations are described by the thermal diffusion equation
coupled to Maxwell's equations

\begin{equation}
\nu\dsf{dT}{dt}=k\dsf{d^2T}{dx^2}+j_cE,
\end{equation}
\begin{equation}
\dsf{d^2E}{dx^2}=\mu\left[\dsf{j_c}{nE_b}\dsf{dE}{dt}-\dsf{dj_c}{dT}\dsf{dT}{dt}\right].
\end{equation}
It should be noted that the nonlinear diffusion-type equations (6)
and (7), totally determine the problem of the space-time
distribution of the temperature and electromagnetic field profiles
in the flux creep regime with a nonlinear current-voltage
characteristics in the semi-infinite sample.

Let us specify the thermal and electrodynamic boundary and initial
conditions to the last system of equations. The thermal boundary
conditions are

$$
\dsf{dT(0, t)}{dx}=0,\quad T(L, t)=T_0,
$$
We assume that the magnetic field perturbation is equal to zero at
the sample surface and according to relation (2), we obtain the
first electrodynamic boundary condition

$$
\dsf{dE(0,t)}{dx}=0.
$$
The second boundary condition for the electric field E(x, t) at
the flux front $x=L$ can be presented as

$$
E(L, t)=0,
$$
The boundary conditions for the magnetic induction are
$$
dB(0, t)=B_e,\quad B(L, t)=0,
$$
where $L=\dsf{cB_e}{4\pi j_c}$  is the London penetration depth.
For initial conditions we assume that the electric field is
uniform within the cross-section of the sample $E=E_0$ at t=0.

Let us derive, for this geometry a differential equations,
describing the spatial and temporal evolution of thermal $T(x, t)$
and electromagnetic field $E(x, t)$ perturbations. We present the
small thermal and electromagnetic perturbations in the form

\begin{equation}
\begin{array}{l}
\Theta(x,t)=T_0(x)+(T_c-T_0)\Theta\exp\left[\gamma
t/t_0+iqz\right],\\
\quad\\
\epsilon(x,t)=E_b(x)+E_b\epsilon\exp\left[\gamma t/t_0+iqz\right].\\
\end{array}
\end{equation}
where $T_0(x)$ and $E_b(x)$ are solutions to the unperturbed
equations obtained in the quasi-stationary approximation
describing the background distributions of temperature and
electric field in the sample. Here $\gamma$ is the eigenvalue of
the problem to be determined and $q=2\pi d/L$ is the wave-number
of the perturbation. From solutions (8), one can see that the
characteristic time of thermal and electromagnetic perturbations
$t$ is of the order of $t_0/\gamma$. Where, we have introduced the
following dimensionless parameters and variables

$$
t_0=\dsf{\sigma\nu a}{j_c},\quad z=\dsf{x}{d},\quad d=\dsf{\nu
a}{\mu_0 j_c},\quad q=\dsf{\pi}{2}\dsf{d}{L}.
$$
As we mentioned above, the background temperature $T_0(x)$ is
practically uniform over the cross-section of the sample and under
this approximation we ignore its coordinate dependence. We note
that the background electric field may be created by ramping the
external magnetic field, and for simplicity we assume it to be
coordinate independent. It turns out that these simplifications
have no qualitative influence on the results but make it possible
to perform analytical calculations completely.

Substituting the expression (8) into the system equations (6), (7)
one can get the following linearized system equations for $\Theta$
and $\epsilon$

\begin{equation}
\begin{array}{l}
\tau
q^2\Theta+\gamma\Theta+\dsf{1}{n}\Theta-2\left(\dsf{1}{n}\right)^2\epsilon=0,\\
\quad\\
q^2\epsilon+\gamma\left[\epsilon-n\Theta\right]=0.\\
\end{array}
\end{equation}
Solving the above system equations (9) we obtain the following
dispersion relation to determine an eigenvalues of the problem

\begin{equation}
\textcolor{black}{\gamma^2+\left[(\tau+1)q^2-\dsf{1}{n}\right]\gamma+\left[\tau
q^2+\dsf{1}{n}\right]q^2=0}
\end{equation}
where $\tau$ is the ratio between the characteristic time of
magnetic flux diffusion and the characteristic time of heat flux
diffusion [4]. The instability of the flux front is defined by the
positive values of the growth rate Re $\gamma$>0. It can be seen
that there is a critical wave number,

\begin{equation}
q_c=\dsf{1}{\sqrt{\tau}}.
\end{equation}
below which the system is always unstable at n=1. This instability
appears first at q=0. In this case the small perturbations grow
with the maximal possible rate Re$\gamma=1$. The growth rate
dependence on the wave number for different values of $\tau$ is
illustrated in Figs. (1-3) at $n=1$. For high enough values of
$\tau$ the system is stable. As the  $\tau$ decreases, the growth
rate $\gamma$ increases. The branching instability will gradually
appears for relatively small values of $\tau$=0.05.

Thus, according to (11), the branching instability occurs at the
threshold electric field $E=E_c$

$$
E=E_c=\dsf{\pi^2}{4}\dsf{\kappa (T_c-T_0)}{j_cL^2}.
$$
Taking into account an expression for penetration depth L, the
threshold field can be written at $B=B_{th}$  as

\begin{equation}
B_{th}=\dsf{\pi}{2}\sqrt{\dsf{\kappa(T_c-T_0)j_c}{E_b}}.
\end{equation}
The threshold field for branching instability, as can be seen from
the last expression is highly sensitive to the critical current
density and the shape of the background electric field $E_b$
generated by the varying magnetic field. The threshold field
$B_{th}$ decreases monotonously with increasing the background
electric field $E_b$. If we assume that the background electric
field $E_b$ generated by a varying magnetic field as
$E_b\simeq\dot{B_e}$, for the considered simple geometry, then we
can easily obtain the expression for the sweep rate dependence of
the threshold field of branching instability.

Let us assume that the thermal diffusion is slower than the
magnetic diffusion $\tau\ll 1$. In this adiabatic limiting case,
the instability criterion is determined as $q=q_c=1$ for n=1, so
the threshold field can be presented as

$$
B_{th}=\dsf{\pi}{2}\sqrt{\dsf{\nu}{\mu_0}(T_c-T_0)}.
$$
This is a well known adiabatic criteria [2], which assumes that
the heat transport from the sample surface to the environment can
be neglected.

\vskip 0.5cm
\begin{center}
{\bf Conclusion}
\end{center}
In conclusion, on the basis of a linear analysis of a set of
differential equations describing small perturbations of
temperature and electromagnetic field we found that under some
conditions a branching instability may occur in the sample.

\vskip 0.5cm
\begin{center}
{\bf   Acknowledgements}
\end{center}
This study was supported by the NATO Reintegration Fellowship
Grant and Volkswagen Foundation Grant. Part of the computational
work herein was carried on in the Condensed Matter Physics at the
Abdus Salam International Centre for Theoretical Physics.

\vskip 0.5cm
\begin{center}
{\bf  References}
\end{center}

\begin{enumerate}
\item C. P. Bean, Phys. Rev. Lett. 8, 250 (1962); Rev. Mod. Phys.,
36, 31 (1964).

\item P. S. Swartz and S.P. Bean, J. Appl. Phys., 39, 4991 (1968).

\item S. L. Wipf, Cryogenics, 31, 936 (1961).

\item  R. G. Mints and A.L. Rakhmanov, Rev. Mod. Phys., 53, 551
(1981).

\item R. G. Mints and A.L. Rakhmanov, Instabilities in
superconductors, Moscow, Nauka, 362 (1984).

\item R. G. Mints, Phys. Rev., B 53, 12311 (1996).

\item R. G. Mints and E.H. Brandt, Phys. Rev., B 54, 12421 (1996).

\item  P. Leiderer, J. Boneberg, P. Brüll, V. Bujok, and S.
Herminghaus, Phys. Rev. Lett. 71, 2646 (1993).

\item U. Bolz, J. Eisenmenger, J. Schiessling, B. U. Runge, and P.
Leiderer, Physica B 284, 757 (2000).

\item U. Bolz, D. Schmidt, B. Biehler, B. U. Runge, R. G. Mints,
K. Numssen, H. Kinder, and P. Leiderer, Physica C 388, 715 (2003).

\item  C. A. Duran, P. L. Gammel, R. E. Miller, and D. J. Bishop,
Phys. Rev. B 52, 75 (1995).

\item  V. Vlasko-Vlasov, U. Welp, V Metlushko, and G. W. Crabtree,
Physica C 341, 1281 (2000).

\item  T. H. Johansen, M. Baziljevich, D. V. Shantsev, P. E. Goa,
Y. M. Galperin, W. N. Kang, H. J. Kim, E. M. Choi, M. S. Kim, and
S. I. Lee, Europhys. Lett. 59, 599 (2002).

\item  T. H. Johansen, M. Baziljevich, D. V. Shantsev, P. E. Goa,
Y. M. Galperin, W. N. Kang, H. J. Kim, E. M. Choi, M. S. Kim, and
S. I. Lee, Supercond. Sci. Technol. 14, 726 (2001).

\item  A. V. Bobyl, D. V. Shantsev, T. H. Johansen, W. N. Kang, H.
J. Kim, E. M. Choi, and S. I. Lee, Appl. Phys. Lett. 80, 4588
(2002).

\item A. L. Rakhmanov, D. V. Shantsev, Y. M. Galperin, and T. H.
Johansen2, Phys. Rev B 70, 224502 (2004).

\item D. V. Shantsev, V. Bobyl, Y. M. Galperin, T. H. Johansen,
and S. I. Lee, Phys. Rev. B, 72, 024541 (2005).

\item  A. M. Campbell and J. E. Evetts, Critical Currents in
Superconductors, (Taylor and Francis, London, 1972), Moscow
(1975).

\item P. W. Anderson , Y.B. Kim  Rev. Mod. Phys., 36 (1964).

\item P. W. Anderson,  Phys. Rev. Lett.,  309, 317 (1962).

\item E. Zeldov, N. M. Amer, G. Koren, A. Gupta, R. J. Gambino,
and M. W. McElfresh, Phys. Rev. Lett., 62, 3093 (1989).

\item P. H. Kes, J. Aarts, J. van der Berg, C. J. van der Beek,
and J.A. Mydosh, Supercond. Sci. Technol., 1, 242 (1989).

\end{enumerate}

}\end{multicols}

\newpage

\begin{center}
\includegraphics[width=3in]{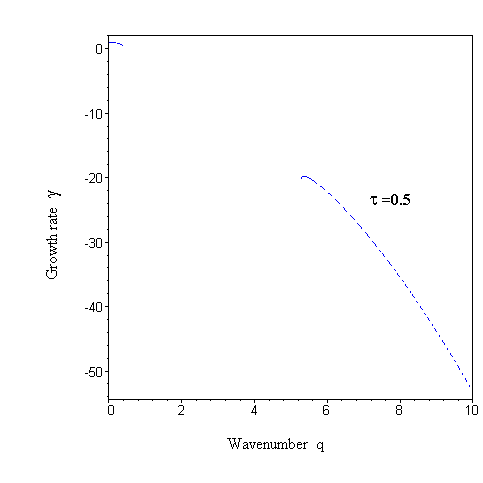}
\includegraphics[width=3in]{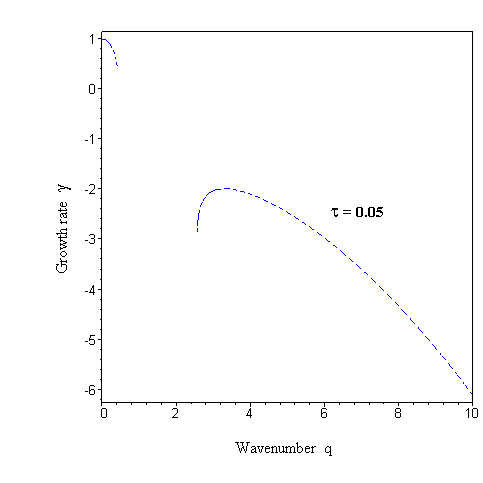}
\includegraphics[width=3in]{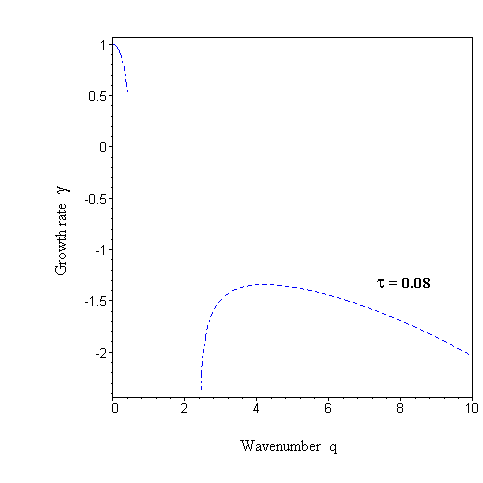}
\end{center}
\begin{center}

Fig.1-3. The dependence of the growth rate on the wave number for
$\tau=0.5, 0.05, 0.08$ and n=1.
\end{center}

\end{document}